\documentclass[twocolumn]{aastex7}

\begin{document}

\title{Evidence for Ambipolar Diffusion from Ion-Neutral Velocity Drift in HH 212}

\shortauthors{L\'opez-V\'azquez et al.}

\author[0000-0002-5845-8722,sname=L\'opez-V\'azquez]{J. A. L\'opez-V\'azquez}
\thanks{Corresponding author: J. A. L\'opez-V\'azquez}
\affiliation{Academia Sinica Institute of Astronomy and Astrophysics, No. 1, Sec. 4, Roosevelt Road, Taipei 10617, Taiwan}
\email[show]{jlopezv@asiaa.sinica.edu.tw}

\author[0000-0002-3024-5864,sname=Lee]{Chin-Fei Lee}
\affiliation{Academia Sinica Institute of Astronomy and Astrophysics, No. 1, Sec. 4, Roosevelt Road, Taipei 10617, Taiwan}
\affiliation{Taiwan Astronomical Research Alliance (TARA), Taiwan}
\email{cflee@asiaa.sinica.edu.tw}

\author[0000-0002-2196-1447,sname=Tapia]{Carlos Tapia}
\affiliation{Facultad de Ciencias de la Tierra y el Espacio, Universidad Auton\'oma de Sinaloa (FACITE-UAS), Josefa Ortiz de Dom\'inguez S/N, C.U. Culiac\'an, Sin. C.P. 80040, M\'exico}
\email{carlostapia@uas.edu.mx}

\author[0009-0006-0486-5468,sname=Lin]{Jyun-Heng Lin}
\affiliation{Institute of Astronomy, National Tsing Hua University, No. 101, Section 2, Kuang-Fu Rd., Hsinchu 30013, Taiwan}
\affiliation{Academia Sinica Institute of Astronomy and Astrophysics, No. 1, Sec. 4, Roosevelt Road, Taipei 10617, Taiwan}
\email{jyunhenglin@asiaa.sinica.edu.tw}

\author[0000-0003-1514-3074,sname=Codella]{Claudio Codella}
\affiliation{INAF, Osservatorio Astrofisico di Arcetri, Largo E. Fermi 5, I-50125 Firenze, Italy}
\email{claudio.codella@inaf.it}

\author[0000-0003-2733-5372]{Linda Podio}
\affiliation{INAF, Osservatorio Astrofisico di Arcetri, Largo E. Fermi 5, I-50125 Firenze, Italy}
\email{linda.podio@inaf.it}

\author[0000-0002-1912-5394,sname=Jaquez]{Jes\'us Miguel J\'aquez-Dom\'inguez}
\affiliation{Instituto de Radioastronom\'ia y Astrof\'isica, Universidad Nacional Aut\'onoma de M\'exico, Antigua Carretera a P\'atzcuaro \#8701, Ex-Hda. San Jos\'e de la Huerta, Morelia, Michoac\'an, C.P. 58089, M\'exico.}
\email{j.jaquez@irya.unam.mx}

\begin{abstract}
We present the Atacama Large Millimeter/submillimeter Array Band 7 observations of the hyperfine structure of C$^{17}$O (3--2) and the H$^{13}$CO$^{+}$ (4--3) molecular line emissions of the Class 0 protostellar system HH 212 at $\sim$ 44 au spatial resolution. Radiative transfer models interpreted in the framework of non-ideal MHD of the envelope allow us to estimate an envelope mass of $M_{env}\sim0.12\,M_\odot$ and a mass infall rate of $\dot{M}_{env}\sim4.4-5.7\times10^{-6}\,M_\odot\,\mathrm{yr}^{-1}$, consistent with values reported in the literature. The models also reveal a systematic radial velocity drift between the neutral tracer (C$^{17}$O (3--2)) and the ionized species (H$^{13}$CO$^{+}$ (4--3)) of $0.1-0.3$ km s$^{-1}$, exceeding the spectral resolution of 0.105 km s$^{-1}$ and consistent with possible ambipolar diffusion in the envelope. The ambipolar diffusion fraction, $f_{AD}\sim 0.1-1$, and the Hall parameter, $\beta_{i,H_2}\gg1$, indicate partial ion–neutral decoupling, with non-ideal effects becoming more significant toward the inner regions. In the envelope, our analysis shows that rotational energy decreases from $\sim 50\,\%$ of the gravitational energy at $R_{cen}$ to $\sim 25\,\%$ at $2R_{cen}$ and $\sim4\,\%$ at 1100 au. In contrast, for $B_0=1$ mG, magnetic energy increases from $\sim 0.4\,\%$ of the gravitational energy at $R_{cen}$ to $\sim2.6\,\%$ at $2R_{cen}$ and $\sim20\,\%$ at 1100 au. The drift velocity expected from radial force balance is consistent with the modeled drift velocity within $\sim300–400$ au of the central source but is lower at larger radii. These results are consistent with a scenario in which ambipolar diffusion leads to partial ion–neutral decoupling, allowing neutrals to move faster than ions within the envelope.
\end{abstract}




\section{Introduction}
\label{sec:introduction}

Low-mass stars form inside molecular clouds through the gravitational collapse of dense protostellar cores with radii $\lesssim0.1$ pc and masses $\sim0.1–10\,M_\odot$ (\citealt{MacLow2004,Pokhrel2018}). Angular momentum and magnetic fields regulate the infall and rotation of the collapsing material, shaping the formation and early evolution of the protostar–disk system (e.g., \citealt{Shu1987,McKee2007,Li2014}). Interferometric observations have revealed that some protostellar systems are surrounded by flattened rotating envelopes consistent with angular momentum conservation (e.g., \citealt{Ohashi1997,Lee2006,Yen2013,Sakai2014,Sai2023}). At these same scales, polarized dust emission, produced by the alignment of non-spherical grains with the local magnetic field (\citealt{Hoang2009,Anderson2015}), has revealed coherent magnetic field structures, showing that magnetic fields are dynamically relevant during collapse and the early stages of protostar–disk formation (e.g., \citealt{Huang2024,Huang2025}).

Ambipolar diffusion (AD) is a non-ideal MHD effect that allows neutral gas to drift relative to ions, redistribute magnetic flux, and reduce the efficiency of magnetic braking. This process facilitates the formation of Keplerian disks larger than 10 au (e.g., \citealt{Dapp2012,Li2014,Tsukamoto2015}) and becomes significant on scales of tens to hundreds of au, where the magnetic field gradually decouples from the neutrals and may form an AD shock that slows infall and rotation (e.g., \citealt{Krasnopolsky2002,Li2011,Zhao2018}). The efficiency of this process depends on magnetic diffusivity, grain properties, and ionization rates, potentially limiting disk growth under standard conditions (\citealt{Mellon2009}). 

Observationally, AD produces measurable differences between the velocities of ionized and neutral tracers in molecular clouds and protostellar envelopes, with neutral tracers showing larger velocity dispersions, as observed in the star-forming molecular clouds M17 (\citealt{Li2008}) and NGC 6334 (\citealt{Tang2018}) from HCN and HCO$^{+}$ observations at scales of $<0.5$ pc. At smaller scales of $\sim$100 au, \citet{Yen2018} compared the ionized H$^{13}$CO$^{+}$(3–2) and neutral C$^{18}$O (2–1) lines in the Class 0 protostar B335, finding no clear ion–neutral velocity drift and constraining it to $\lesssim0.3$ km s$^{-1}$, suggesting that the magnetic field remains well coupled to the neutral material at these radii. Theoretically, \citet{Lankhaar2020} proposed that ion–neutral drift from AD can partially align molecular ions, producing collisional polarization with vectors perpendicular to the magnetic field, while \citet{Thieme2024} developed an analytical framework to estimate ambipolar diffusivity from observable disk properties, highlighting its dynamical importance in protostellar disks.

Located in the Orion-L1630 cloud at a distance of $\sim423\pm21$ pc (\citealt{Zucker2019}), the Class 0 protostellar system HH 212 has an age of $\sim40000$ yr (\citealt{Lee2014}) and a central source mass of $0.25\pm0.05\,M_\odot$ (\citealt{Codella2014,Lee2014}). HH 212 hosts a dusty edge-on accretion disk (\citealt{Lee2017Sc}) with an inclination angle with respect to the line of sight of $i\sim86^{\circ}$ (\citealt{Claussen1998}). Observations of C$^{18}$O (2-1) and $^{13}$CO (2-1) obtained with the Submillimeter Array (\citealt{Lee2006}), as well as C$^{17}$O and HCO$^{+}$ observations obtained with the Atacama Large Millimeter/submillimeter Array (ALMA) (\citealt{Codella2014,Lee2014,Lee2017ApJ}), have revealed that the system has a rotating and flattened envelope around the central source with rotation consistent with angular momentum conservation, corresponding to $\sim140$ au km s$^{-1}$ (\citealt{Lee2017ApJ}). In this paper, we present ALMA observations of the C$^{17}$O (3--2) and H$^{13}$CO$^{+}$ (4--3) molecular line emissions. This work is organized as follows: Section \ref{sec:observations} describes the observations, Section \ref{sec:results} presents the results, Section \ref{sec:discussion} discusses them, and Section \ref{sec:conclusions} summarizes the conclusions.

\begin{figure*}[t!]
\centering
\includegraphics[scale=0.4]{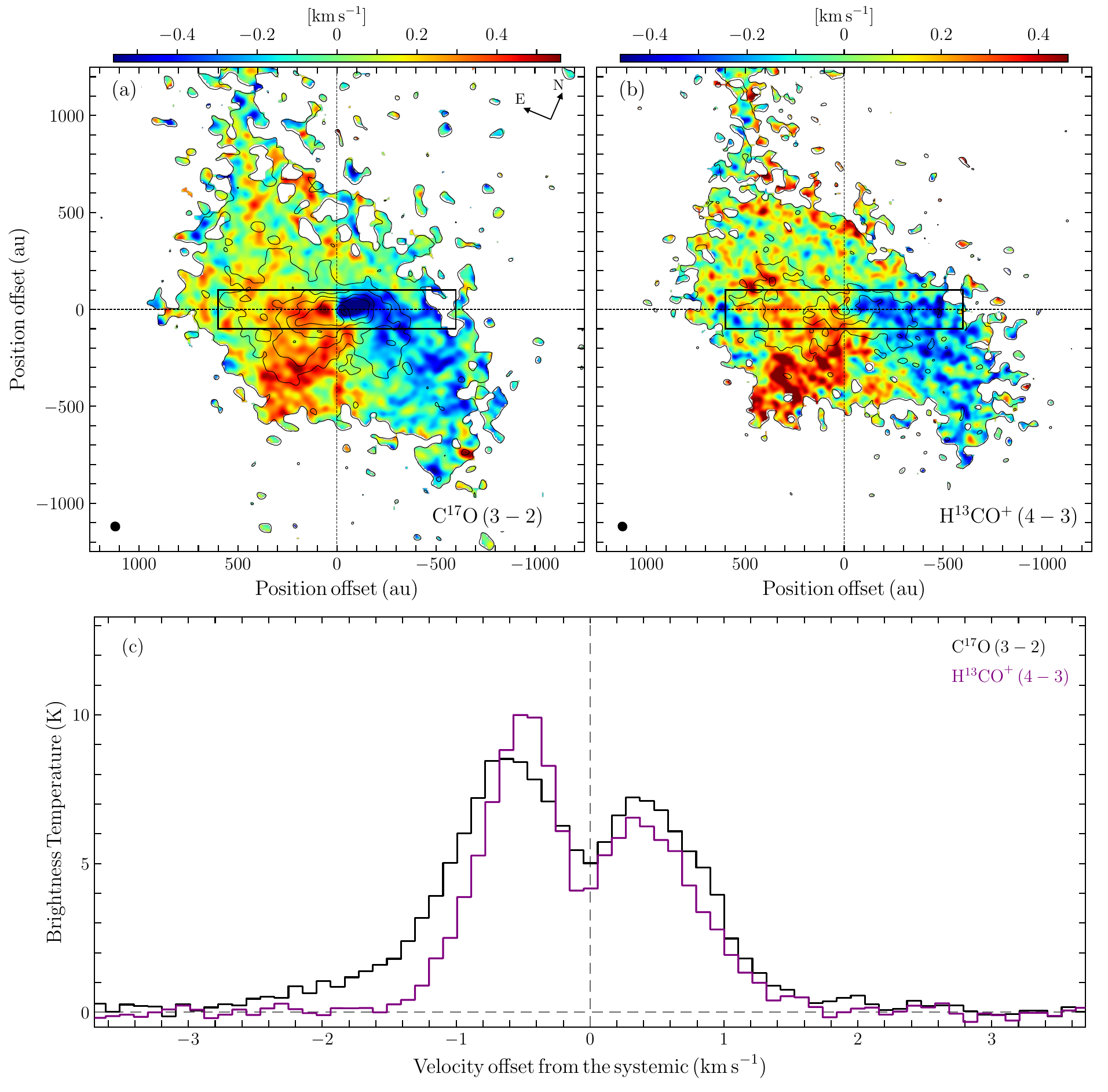}
\caption{HH 212 envelope emission of the C$^{17}$O (3--2) and H$^{13}$CO$^{+}$ (4--3)  molecular lines. (a) ALMA first-moment map, or the intensity-weighted velocity, of C$^{17}$O (3--2). (b) ALMA first-moment map of H$^{13}$CO$^{+}$ (4--3). (c) Spectra averaged over a rectangular region of $3^{\prime\prime}.0\times0^{\prime\prime}.5$ centered at the central source, aligned with the jet axis. Color codes are consistent across all labels. The dashed vertical and horizontal gray lines mark the systemic velocity and the zero brightness temperature, respectively. The contour levels in panels (a) and (b) correspond to ALMA moment-zero maps, or integrated intensity, starting at 3$\sigma$ and increasing in steps of 2$\sigma$ up to 11$\sigma$, where $\sigma=3.4$ and 3.9 mJy beam$^{-1}$ km s$^{-1}$ for C$^{17}$O (3--2) and H$^{13}$CO$^{+}$ (4--3), respectively. The synthesized beams are shown in the lower-left corner of panels (a) and (b).}
\label{fig:moments_obs}
\end{figure*}

\section{Observations}
\label{sec:observations}

ALMA Band 7 observations of HH 212 were carried out on October 6 and November 26, 2016 (program 2016.1.01475.S, PI: Claudio Codella) during Cycle 4. In this work, we focus on the emission from the hyperfine structure of C$^{17}$O (3--2) line and the H$^{13}$CO$^{+}$ (4--3) line. These observations have been described in detail in \citet{Bianchi2017} and \citet{Tabone2017}. The data were calibrated using the Common Astronomy Software Application (CASA) package (\citealt{CASAteam}), version 4.7.0. During the imaging process, Briggs weighting with a robust parameter of 0.5 was applied to balance sensitivity and angular resolution.

The rms noise levels for C$^{17}$O (3--2) and H$^{13}$CO$^{+}$ (4--3) are 2.6 and 2.8 mJy beam$^{-1}$, respectively. Both datasets have a spectral resolution of 0.105 km s$^{-1}$, and their synthesized beams are $0^{\prime\prime}.11\times 0^{\prime\prime}.11$ with position angles of $-56^\circ.7$ for C$^{17}$O and $-55^\circ.6$ for H$^{13}$CO$^{+}$. 

\section{Results}
\label{sec:results}

To facilitate the interpretation of maps and figures, we define an offset velocity $V_{off} = V_{lsr}-V_{sys}$, where the systemic velocity of HH 212 is $V_{sys} = 1.7\pm0.1$ km s$^{-1}$ (\citealt{Lee2014}). All maps are rotated by $22^\circ.5$ clockwise so that the jet axis aligns with the $y-$axis.

\begin{figure*}[t!]
\centering
\includegraphics[width=\linewidth]{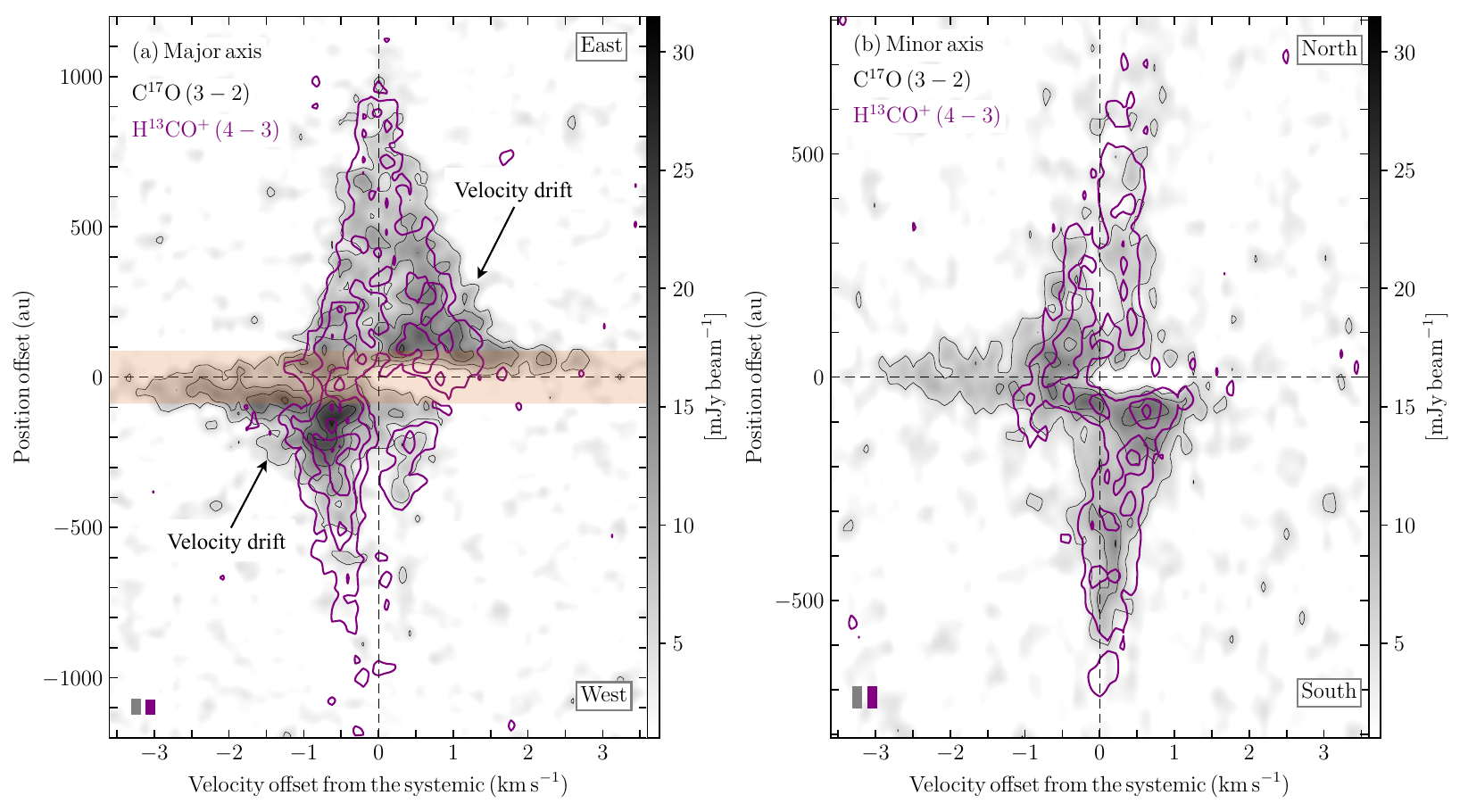}
\caption{PV diagrams of the C$^{17}$O (3--2) (grayscale) and H$^{13}$CO$^{+}$ (4--3) (purple contours) molecular line emission from the envelope, along (a) the major axis and (b) the minor axis. The shaded region in panel (a) represents the area inside the centrifugal radius. The black arrows indicate the velocity drift between the neutral (C$^{17}$O) and the ionized (H$^{13}$CO$^{+}$) species. In both panels, the contour levels start at 2$\sigma$ and increase in steps of 2$\sigma$ up to 10$\sigma$, where $\sigma=2.6$ and 2.8 mJy beam$^{-1}$ for C$^{17}$O (3--2) and H$^{13}$CO$^{+}$ (4--3), respectively. The gray and purple bars in the lower-left corner indicate the channel width and the angular resolution.}
\label{fig:pvobservations}
\end{figure*}

\subsection{Infalling-Rotating Envelope}
\label{subsec:Infalling_envelope}

Figure \ref{fig:moments_obs} shows the ALMA first-moment (intensity-weighted velocity) maps of the C$^{17}$O and H$^{13}$CO$^{+}$ (panels \ref{fig:moments_obs}a and b), with total-intensity maps overlaid in black contours. The maps were integrated over $\pm3$ km s$^{-1}$ relative to the systemic velocity. Both molecules trace the extended envelope out to $R\sim\pm2^{\prime\prime} \approx\pm800$ au and show blueshifted velocities to the west and redshifted velocities to the east, consistent with rotation (\citealt{Codella2014,Lee2014,Lee2017ApJ}).

Figure \ref{fig:moments_obs}c presents the C$^{17}$O (black line) and H$^{13}$CO$^{+}$ (purple line) spectra extracted from the central region of the envelope, defined by a $3^{\prime\prime}.0\times0^{\prime\prime}.5$ box. H$^{13}$CO$^{+}$ traces only the low-velocity components ($\sim\pm1.5$ km s$^{-1}$), while C$^{17}$O highlights both the low- and high-velocity components ($\sim\pm2.5$ km s$^{-1}$). This pattern is also observed in the position-velocity (PV) diagram along the major axis (perpendicular to the jet axis) presented in Figure \ref{fig:pvobservations}a, where H$^{13}$CO$^{+}$ emission (purple contours) within a radius of $R\le 300$ au shows only velocities up to $\sim\pm1.2$ km s$^{-1}$, while the C$^{17}$O (grayscale) reveals higher-velocity components extending to $\sim\pm1.7$ km s$^{-1}$. The difference between H$^{13}$CO$^{+}$ and C$^{17}$O suggests a lower-density innermost region, with C$^{17}$O tracing this more tenuous gas due to its lower critical density, in contrast to H$^{13}$CO$^{+}$. This behavior is similar to that observed in IRAS 15398-3359 (\citealt{Jorgensen2013}). In addition, C$^{17}$O is also present in the denser disk (\citealt{Tabone2017}), potentially contaminating the emission. At the same time, the slightly higher velocities observed for C$^{17}$O in the innermost regions may indicate a small velocity drift between the two tracers (black arrows in Figure \ref{fig:pvobservations}a), reflecting slight differences in the gas motions within the same physical region without implying distinct velocity components. This velocity drift is clearly detected, being consistently observed across the sampled envelope and exceeding the spectral resolution of 0.105 km s$^{-1}$.

PV diagram along the minor axis of the envelope (aligned with the jet axis) is shown in Figure \ref{fig:pvobservations}b. The emission within $\pm 1$ km s$^{-1}$ is dominated by the envelope around the central source, based on the spatial and kinematic distributions of the C$^{17}$O and H$^{13}$CO$^{+}$. The high-velocity C$^{17}$O emission near the position zero traces the disk (\citealt{Tabone2017}).

\begin{figure*}[t!]
\centering
\includegraphics[width=\linewidth]{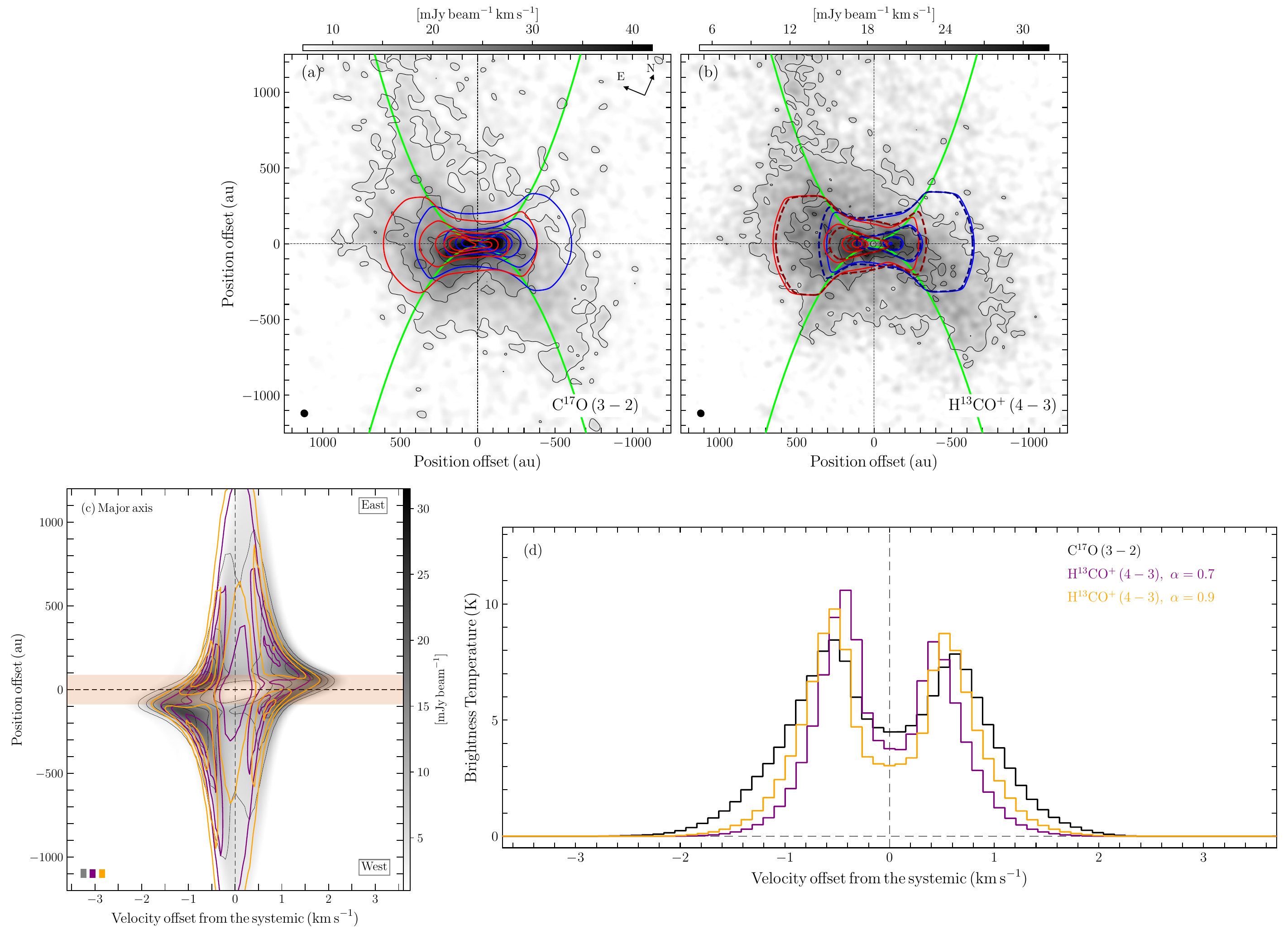}
\caption{Comparison between the observed emission of the envelope and the corresponding best-fit models. Panels (a) and (b) show the ALMA moment-zero or integrated intensity maps of C$^{17}$O (3--2) and H$^{13}$CO$^{+}$ (4--3), respectively, overlaid with the blueshifted and redshifted model contours. In panel (b) the solid lines correspond to the model with $\alpha=0.7$ while the dashed lines respresent the model with $\alpha=0.9$. Panel (c) presents the PV diagram of the best-fit models of both lines together, with the shaded region marking the area inside the centrifugal radius. Panel (d) displays the spectra of both lines as in Figure \ref{fig:moments_obs}c, with the H$^{13}$CO$^{+}$ (4-3) model spectrum shown after subtracting a Gaussian component centered at 0.1 km s$^{-1}$. In panels (a) and (b), green parabolas represent the cavity walls of the outflow and the black contour levels follow the same observed integrated-intensity levels described in Figure \ref{fig:moments_obs}, while blue and red contours start at 3$\sigma$ and increase in steps of 2$\sigma$ up to 11$\sigma$, where $\sigma=1.8$ and 1.9 mJy beam$^{-1}$ km s$^{-1}$ for C$^{17}$O (3--2) and H$^{13}$CO$^{+}$ (4--3), respectively. In panel (c) the contours and the bars are the same as described in Figure \ref{fig:pvobservations}. Color codes in panels (c) and (d) are consistent across all labels. The synthesized beams are shown in the lower-left corner in panels (a) and (b).}
\label{fig:moment_models}
\end{figure*}

\subsection{The Infall Model}
\label{subsec:ulrichmodel}

To simultaneously reproduce the emission, kinematics, and morphology of both molecular lines observed in the infalling-rotating envelope of HH 212, we adopt a model that includes both rotational and collapse motions. In this framework, infall follows the gravitational collapse of a rotating molecular cloud, as formulated by \citet{Ulrich1976}. This model assumes that all particles, initially at rest, rotate and fall toward the central protostar while conserving energy and specific angular momentum. If the particles move supersonically during the collapse and heating by radiation and viscosity is negligible, the collapse is pressureless; as a result, the particle orbits are ballistic.

Ulrich's model assumes that both the velocity field and the density profile are axisymmetric with respect to the azimuthal coordinate $\phi$. Under this assumption, the velocity field in spherical coordinates ($r$, $\theta$, and $\phi$) can be expressed as
\begin{equation}
    \label{eq:vrad}
    V_{r} = -V_0 \left(\frac{R_{cen}}{r}\right)^{1/2}\left(1+\frac{\cos\theta}{\cos\theta_0}\right)^{1/2},
\end{equation}

\begin{equation}
    \label{eq:vtheta}
    V_\theta = V_0\left(\frac{R_{cen}}{r}\right)^{1/2}\left(\frac{\cos\theta_0-\cos\theta}{\sin\theta}\right)\left(1+\frac{\cos\theta}{\cos\theta_0}\right)^{1/2},
\end{equation}
and
\begin{equation}
    \label{eq:vphi}
    V_\phi=V_0\left(\frac{R_{cen}}{r}\right)^{1/2}\left(\frac{\sin\theta_0}{\sin\theta}\right)\left(1-\frac{\cos\theta}{\cos\theta_0}\right)^{1/2},
\end{equation}
where $\theta_0$ is the initial polar angle of the orbit of a fluid element at the start of its collapse toward the center. The Keplerian velocity, $V_0$, at the centrifugal radius $R_{cen}$, is given by
\begin{eqnarray}
    \label{eq:velkep}
    V_0 = \left(\frac{G M_*}{R_{cen}}\right)^{1/2},
\end{eqnarray}
where $G$ is the gravitational constant and $M_*$ is the stellar mass.

Since $\rho = 1.4 n m_{H_2}$, where $m_{H_2}$ is the hydrogen molecular mass, the density profile can be equivalently expressed in terms of the mass density $\rho$ or number density $n$, as in the Ulrich solution below
\begin{equation}
    \label{eq:density}
    n=-n_0 \left(\frac{V_0}{V_r}\right)\left(\frac{R_{cen}}{r}\right)^2\left[1+2\left(\frac{R_{cen}}{r}\right)P_2(\cos\theta_0)\right]^{-1}.
\end{equation}
Here, the Legendre polynomial is $P_2(\cos\theta_0)=\frac{1}{2}\left(3\cos\theta_0^2-1\right)$. The angle $\theta_0$ is given implicitly in terms of variables $\theta$, $r$, and $R_{cen}$ by
\begin{equation}
    \label{eq:theta0}
    \frac{R_{cen}}{r}=\frac{\cos\theta_0-\cos\theta}{\sin^2\theta_0\cos\theta_0}.
\end{equation}
Equation (13) of \citet{Mendoza2004} gives an explicit solution to this equation.

The temperature of the envelope is described by a power-law profile,
\begin{equation}
    \label{eq:temperature}
    T = T_0\left(\frac{r}{R_{0}}\right)^{-q},
\end{equation}
consistent with radiative equilibrium in centrally heated spherical envelopes, where $T_0$ denotes the temperature at the reference radius $R_0$ and $q$ is the power-law index.

The Ulrich envelope is described by the stellar mass $M_*$, centrifugal radius $R_{cen}$, and inclination angle $i$. For HH 212, we adopt $M_*=0.25\pm0.05\,M_\odot$ (\citealt{Codella2014,Lee2014}), $R_{cen}=88$ au (\citealt{Lee2017ApJ}), and $i=86^{\circ}$ (\citealt{Claussen1998}). Radiative transfer was performed using RADMC-3D (\citealt{radmc2012}) under LTE. The envelope emission is assumed to extend up to an outer radius of $R_{outer}\sim 2000$ au, as determined from C$^{18}$O observations (\citealt{Lee2006}), and its geometry is that of a flattened envelope, where the half-thickness $H$ is described by (\citealt{Lee2010})
\begin{equation}
    \label{eq:zthicknes}
    H(R) = \mathrm{max}(H_0,R\tan\theta_{env})
\end{equation}
where $R=r\sin\theta$ is the cylindrical radius, $H_0$ is the minimum value of the half-thickness near the source, and $\theta_{env}$\footnote{The notation was changed from $\theta_0$ to $\theta_{env}$ to avoid confusion with the Ulrich envelope} is the flaring angle measured from the midplane (see Figure 1 of \citealt{Lin2025}). The emission from within the outflow-cavity walls, delimited by the C$^{34}$S molecular line emission and fitted by a parabolic shape $z=R^2/a+0^{\prime\prime}.05$ with $a = 0^{\prime\prime}.9$ in the north and $a = 1^{\prime\prime}$ in the south (\citealt{Tabone2017}), is excluded to isolate the envelope contribution. Additionally, disk emission is not considered due to insufficient angular resolution. Therefore, no emission is considered within the centrifugal radius ($R_{cen}$) of $\sim88$ au (\citealt{Lee2017ApJ}).

\begin{figure*}[t!]
\centering
\includegraphics[scale=0.55]{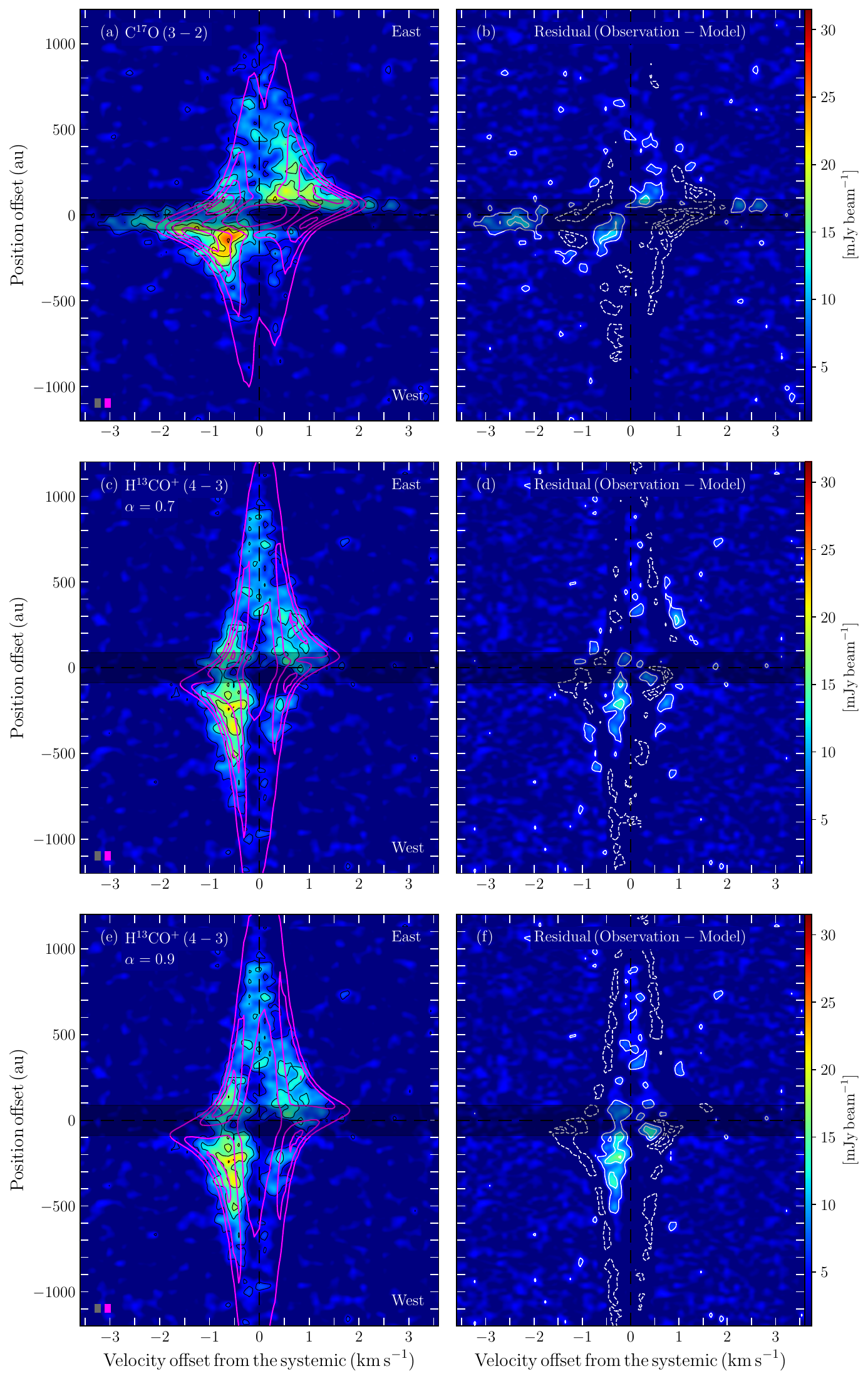}
\caption{The first column shows PV diagrams along the major axis, comparing the observations (black contours and color scale) and the models (magenta contours), and the second column displays the residual maps of the emission of C$^{17}$O (3--2) (top row) and H$^{13}$CO$^{+}$ (4--3) (middle and bottom rows). The shaded region, contour levels, and bars are the same as described in Figure \ref{fig:pvobservations}.}
\label{fig:pvmodels}
\end{figure*}

To compare our observations with the model, we introduce a parameter $\alpha$, such that $V_{r,obs} = \alpha V_r$, because the radial velocity from the Ulrich model (eq. \ref{eq:vrad}) is too high. Although the model is not fully self-consistent with $\alpha$, it still provides a useful estimate of the velocity trends when infall is suppressed (\citealt{Sai2022}). We emphasize that the Ulrich model describes test-particle trajectories in the gravitational potential of a point mass and does not include magnetic fields and pressure gradients. Therefore, the parameter $\alpha$ should not be interpreted as encoding a specific physical mechanism, but rather as a proxy for departures from the idealized assumptions of the model. In a magnetized envelope, such departures from the purely gravitational solution may in principle be associated with additional forces, including magnetic forces. However, because the Ulrich model does not include magnetic forces, the parameter $\alpha$ alone cannot be used to determine the magnetic field strength. The possible contribution of magnetic forces is instead assessed separately by considering the corresponding radial force balance.

To reproduce the emission of the two observed molecular lines, the radiative transfer model searches for the best-fitting combination of six free parameters describing the radial velocity ($\alpha$), the density and temperature ($n_0$, $T_0$, $q$, and $R_0$), and the molecular abundances relative to H$_2$ ($X_{mol/H_2}$). The best-fit parameters for both C$^{17}$O and H$^{13}$CO$^{+}$ correspond to the same density and temperature structure, with $n_0\sim 4\times10^7$ cm$^{-3}$, $T_0\sim 45$ K, $q=0.4$, and $R_0= 176$ au (corresponding to $2R_{cen}$). To ensure a consistent comparison between tracers and minimize degeneracies in the modeling, these physical parameters are fixed to the same values in both cases. In this framework, only the radial velocity parameter $\alpha$ and the molecular abundances are allowed to vary between the two tracers. 

For C$^{17}$O, the abundance is described by a drop profile motivated by CO freeze-out and sublimation in protostellar envelopes (\citealt{Jorgensen2004b,Jorgensen2005}). We adopt an abundance of $X_{C^{17}O/H_2}=2\times10^{-8}$ in regions with temperatures above 20 K, while at lower temperatures the abundance is reduced by a factor of 100 to account for CO depletion onto dust grains. Such abundance variations are expected as CO freezes out in the cold outer envelope and returns to the gas phase once the temperature exceeds its sublimation threshold. The adopted inner abundance is lower than the values reported in the literature of $\sim 4-5\times10^{-8}$ (\citealt{Frerking1982,Ladd2004,Wistrom2006}). For this tracer, we adopt a parameter $\alpha=0.9$.

For H$^{13}$CO$^{+}$, the abundance is not treated as a free parameter. Instead, it is computed using the chemical scheme described by \citet{Aikawa2015} and \citet{Booth2019}, in which the HCO$^+$ abundance is determined by the CO, H$_3^+$, and electron abundances, and the ionization rate $\zeta$ of the envelope. The CO abundance relative to H$_2$ is assumed to be $10^{-4}$ for temperatures above 20 K and depleted by a factor of 100 below this threshold to account for CO freeze-out onto dust grains, following the same prescription adopted for C$^{17}$O. The H$^{13}$CO$^{+}$ abundance is obtained assuming a $^{12}$C/$^{13}$C isotopic ratio (see section 5 of \citealt{Booth2019}). Two parameter sets are explored: one with $\alpha=0.7$, $^{12}$C/$^{13}$C = 60, in agreement with the value of $\sim60$ reported for the same source by \citet{FernandezRuiz2026}, and $\zeta=1\times10^{-16}$ s$^{-1}$, an intermediate value between that measured in Orion B by \citet{Bergin1999} and that reported for this source by \citet{FernandezRuiz2026}; and another with $\alpha=0.9$ (the same as C$^{17}$O), $^{12}$C/$^{13}$C = 20, consistent with measurements toward Class 0/I objects in Perseus \citealt{Busch2025}, and $\zeta=1\times10^{-17}$ s$^{-1}$.

For both lines we adopt a half-thickness of the inner region $H_0\sim0^{\prime\prime}.8$, estimated from HCO$^{+}$ and continuum observations in \citet{Lee2014}. For C$^{17}$O, the hyperfine structure of the $J=3-2$ transition is included, considering the seven strongest $F$ components ($F= 5/2-5/2$, $9/2-7/2$, $3/2-3/2$, $7/2-7/2$, $7/2-5/2$, $5/2-3/2$, and $9/2-9/2$), and we confirm that the broader linewidth is not caused by the hyperfine structure. Best-fit parameters were determined through a simultaneous comparison between the models and observations, considering the channel maps, the integrated spectra, and the PV diagram along the major axis, as depicted in Figure \ref{fig:moment_models}. 

As shown in Figures \ref{fig:moment_models}a and b, the models reproduce the flattened envelope traced by both lines quite well; however, the models for both molecules cannot reproduce the emission observed around the outflow-cavity walls (green parabolas) because this emission is likely associated with material entrained by the outflow, which is not included in our models. The PV diagrams and spectra presented in Figures \ref{fig:moment_models}c and d closely resemble those observed (see Figures \ref{fig:moments_obs}c and \ref{fig:pvobservations}a), with C$^{17}$O tracing both the low- and high-velocity components, while H$^{13}$CO$^+$ displays only the low-velocity components.

\begin{figure*}[t!]
\centering
\includegraphics[scale=0.5]{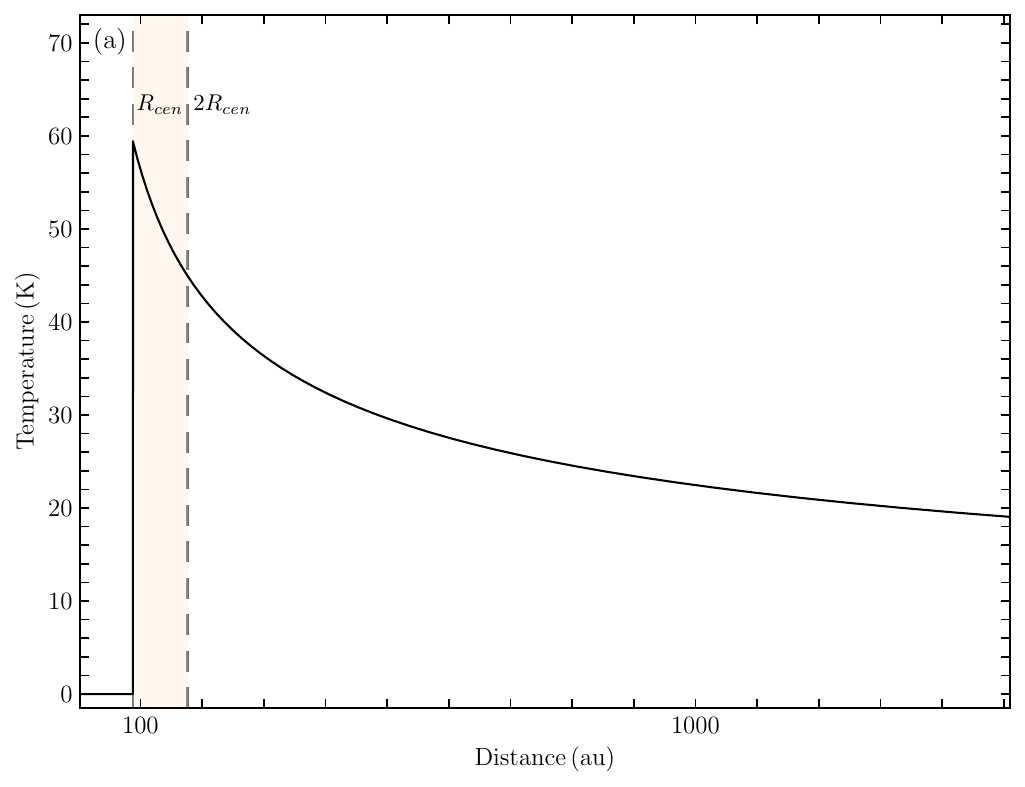}\includegraphics[scale=0.5]{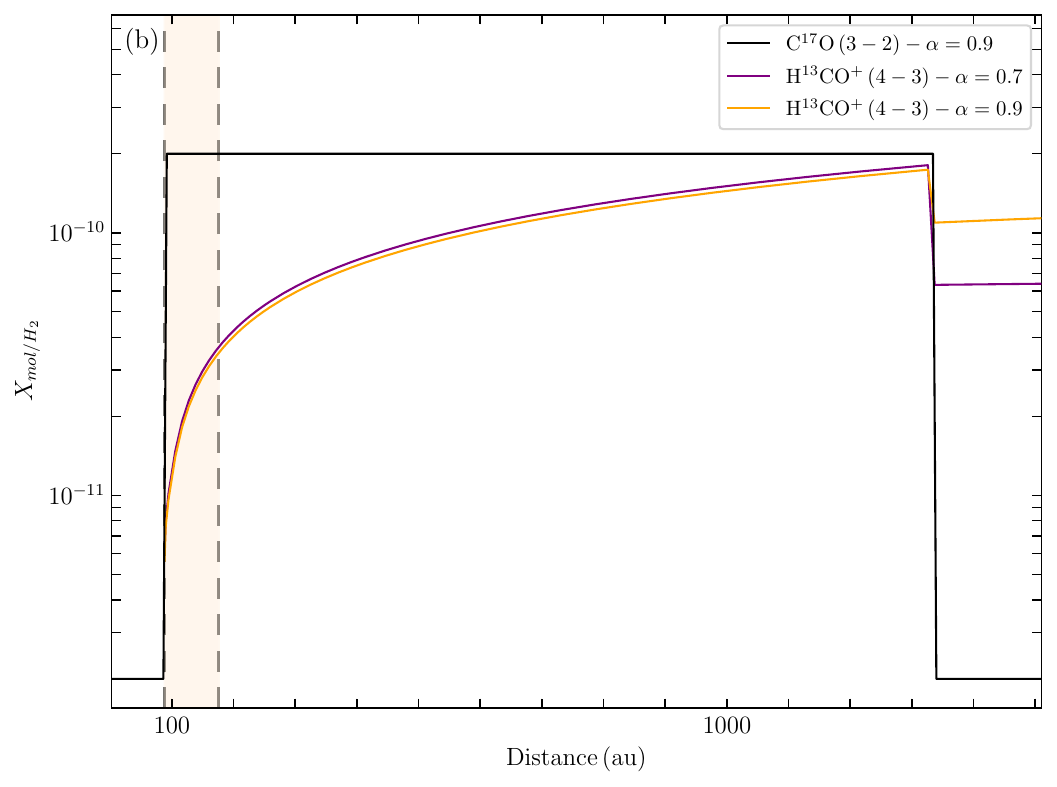}
\caption{Radial profiles of temperature and molecular abundance adopted for the HH 212 envelope. (a) Temperature. (b) Abundance profiles of C$^{17}$O and H$^{13}$CO$^{+}$. For visualization purposes, the C$^{17}$O abundance has been divided by $10^{2}$. The shaded region indicates the radial range between $R_{\rm cen}$ ($\sim88\,\mathrm{au}$) and $2R_{\rm cen}$ ($\sim176\,\mathrm{au}$).}
\label{fig:tempabun}
\end{figure*}

The comparison between the observed and modeled PV diagrams, together with the respective residuals of both molecular lines, is shown in Figure \ref{fig:pvmodels}. The model of C$^{17}$O cannot reproduce the high-velocity components near the central position because the disk emission is not included in our models.

\subsection{Envelope Mass and Infall Rate}
\label{subsec:mass}

We estimated the mass of the envelope using  H$^{13}$CO$^{+}$ line emission, which is given by
\begin{eqnarray}
    \label{eq:massenv}
    M_{env}&=&1.4 m_{H_2}\int_{2R_{cen}}^{R_{outer}} \int_{0}^{2\theta_{env}} \,\left(2\pi\right)\, r\, \left(2H(R)\right)\, n \,d\theta \,dr \nonumber \\
    &\sim& 0.12\,M_\odot,
\end{eqnarray}
which is consistent with previous estimates by \citet{Lee2014} of $\sim0.1\,M_\odot$.

The mass infall rate from the envelope to the protostar-disk system, assuming the infall velocity as in eq. (\ref{eq:vrad}), is
\begin{eqnarray}
    \label{eq:dotmass}
    \dot{M}_{env} &=& 1.4 m_{H_2} \,\left(2R_{cen}\right)\, \left(2\pi\right)\,\int_{0}^{2H(R)}\int_{0}^{2\theta_{env}} z n V_{r,obs} d\theta dz\nonumber \\
    &\sim&4.4-5.7\times10^{-6}\,M_{\odot}\,\mathrm{yr}^{-1},
\end{eqnarray}
where the range reflects the two H$^{13}$CO$^{+}$ models, which adopt different values of the parameter $\alpha$ and therefore different infall velocities. These values are broadly consistent with the estimate reported by \citet{Lee2014} of $\sim5\times10^{-6}\,M_{\odot}\,\mathrm{yr}^{-1}$.

\section{Discussion}
\label{sec:discussion}

To match the observations with the Ulrich ballistic model, we assume that the infall velocity is a fraction $\alpha$ of the radial velocity. For the neutral tracer (C$^{17}$O), we adopt $\alpha=0.9$, while for the ionized tracer (H$^{13}$CO$^{+}$) we use two different values, $\alpha=0.9$ and $\alpha=0.7$. The former corresponds to a model with a low ionization rate ($\zeta=1\times10^{-17}$ s$^{-1}$) and a lower $^{12}$CO/$^{13}$CO isotopic ratio of 20, while the latter assumes a higher ionization rate ($\zeta=1\times10^{-16}$ s$^{-1}$) and a higher isotopic ratio of 60.

In the case where both tracers are described by the same fraction of the radial velocity ($\alpha=0.9$), the velocity drift observed between C$^{17}$O and H$^{13}$CO$^{+}$ may be explained by variations in molecular abundances and the chemical structure of the envelope. In this scenario, abundance gradients and depletion effects can reproduce the observed velocity differences between two tracers.

In contrast, for models with different values of $\alpha$ for the two species, the observed velocity drift may reflect a partial dynamical decoupling between ionized and neutral gas components. In this case, ions and neutrals do not necessarily follow identical infall motions, which is consistent with effects like those produced by AD in the envelope.

\subsection{Abundance Variations}
\label{subsec:abundances}

The observed velocity drift between the neutral tracer (C$^{17}$O) and the ionized tracer (H$^{13}$CO$^{+}$) can be explained by variations in the abundances of both species throughout the envelope. In our models, the abundance of CO, and consequently that of C$^{17}$O, is primarily regulated by the gas temperature, accounting for CO freeze-out onto dust grains in the colder regions (e.g., \citealt{Jorgensen2013}). For the HH 212 envelope, we assume a CO sublimation temperature of $\sim20$ K. In the envelope midplane, this temperature is reached at a radius of $\sim1300$ au from the central source (Figure \ref{fig:tempabun}a). At smaller radii, where CO is returned to the gas phase, the abundance of CO, and consequently that of C$^{17}$O, increases by two orders of magnitude (black line in Figure \ref{fig:tempabun}b). Since the formation of H$^{13}$CO$^{+}$ depends on the availability of gaseous CO (e.g., \citealt{Aikawa2015}), its abundance also exhibits a radial variation (purple and orange lines in Figure \ref{fig:tempabun}b). As a result, the two tracers may probe slightly different regions of the envelope, which could contribute to the observed velocity drift.

Although this chemical scenario reproduces the observed velocity drift (Figure \ref{fig:moment_models}d), the H$^{13}$CO$^{+}$ model adopting the same fraction of the radial velocity ($\alpha=0.9$) as C$^{17}$O exhibits a systematic excess of emission in both the blueshifted and redshifted components at large distances from the central source, as shown by the residual map in Figure \ref{fig:pvmodels}f, indicating a less satisfactory overall fit than the model with $\alpha=0.7$ (Figure \ref{fig:pvmodels}d).

Chemical differentiation and radiative-transfer effects can, in principle, produce apparent ion–neutral velocity differences (\citealt{Tritsis2023}). However, the use of optically thin isotopologues (H$^{13}$CO$^{+}$ and C$^{17}$O) reduces biases associated with optical depth and provides a more reliable probe of the envelope structure. Although these species are not expected to strictly co-evolve (e.g., \citealt{Tassis2012}), they trace broadly overlapping density regimes in HH 212, allowing a consistent comparison of the regions probed by each tracer. In our models, the H$^{13}$CO$^{+}$ abundances do not show significant variations between the cases considered (purple and orange curves in Figure \ref{fig:tempabun}b), and the adopted values of the isotopic ratio and ionization rate in the model with $\alpha=0.7$ are consistent with previous estimates for low-mass protostellar systems such as HH 212. In this context, since abundance variations do not appear to dominate the differences between models, the observed velocity drift in HH 212 is more consistently interpreted as a transition in the velocity field between the neutral and ionized gas, rather than as a purely chemical effect.

\begin{figure*}[t!]
\centering
\includegraphics[scale=0.5]{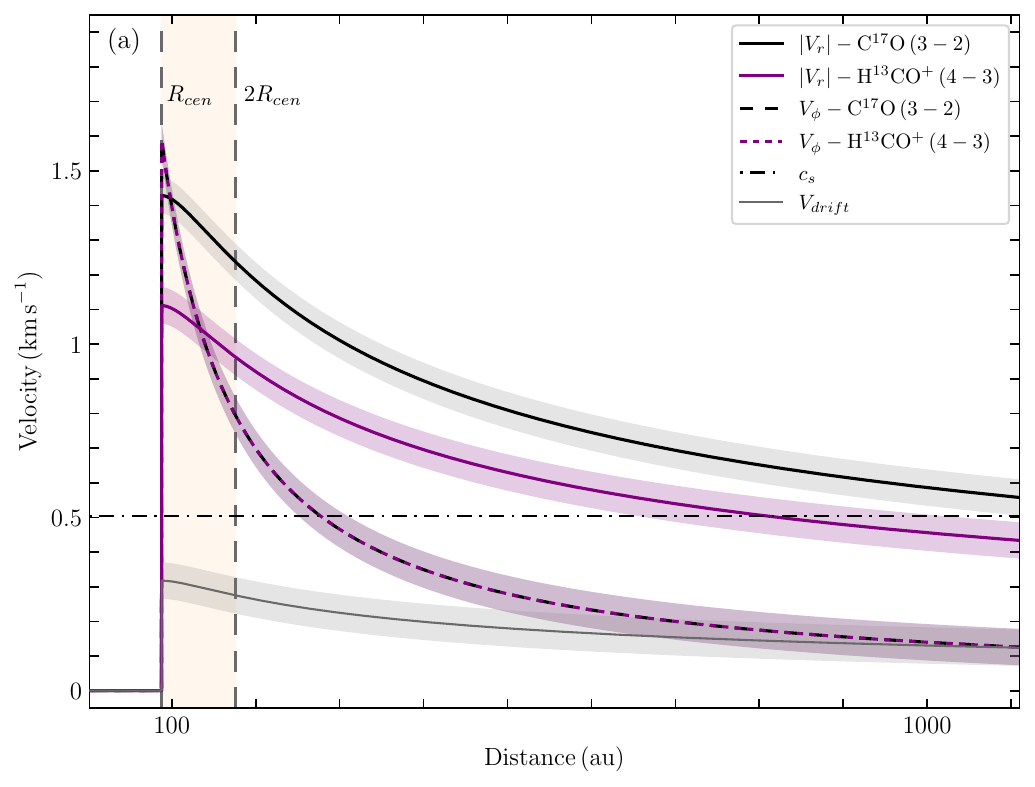}\includegraphics[scale=0.5]{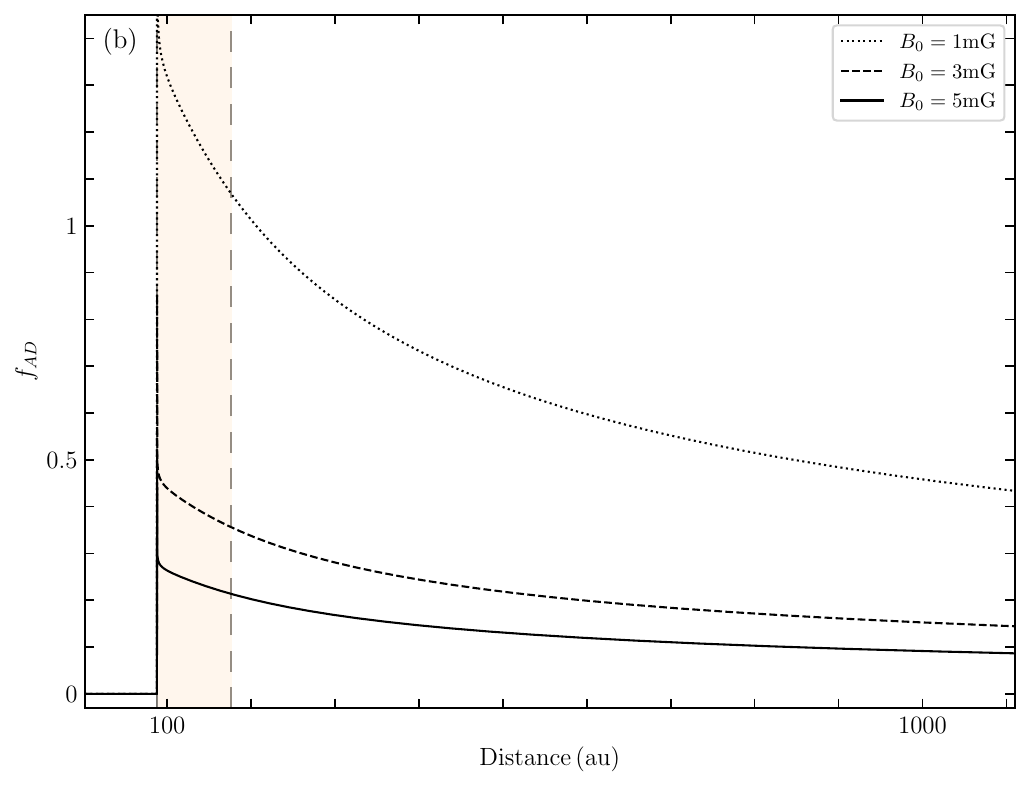}
\includegraphics[scale=0.5]{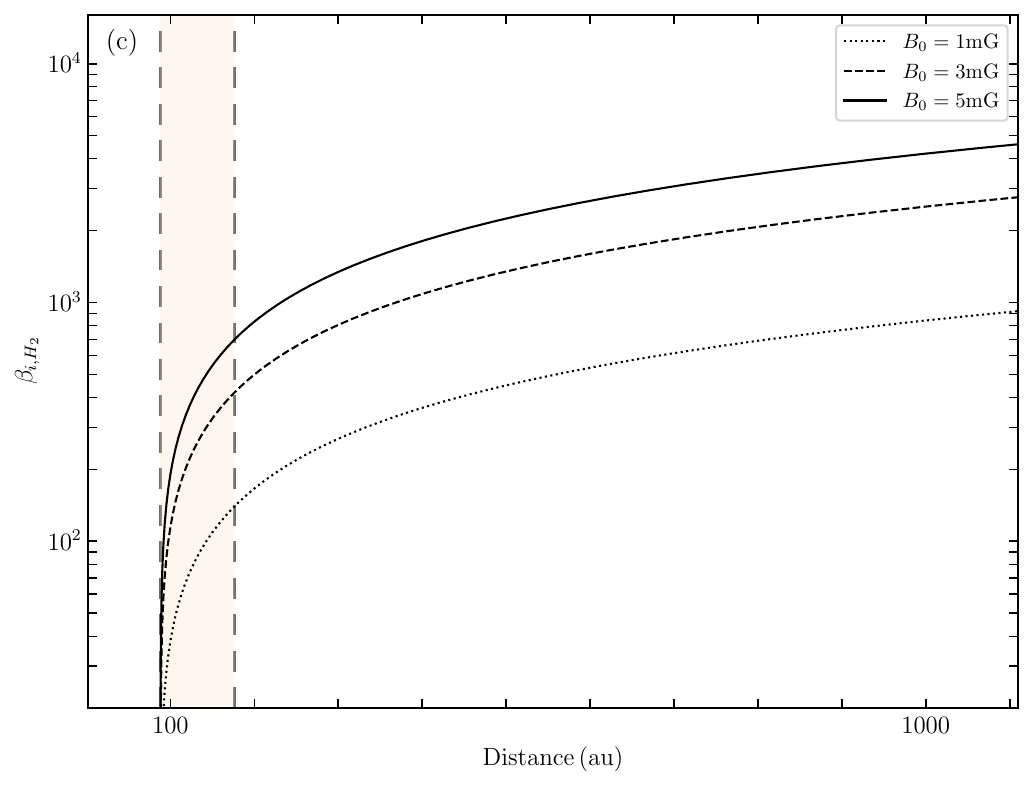}\includegraphics[scale=0.5]{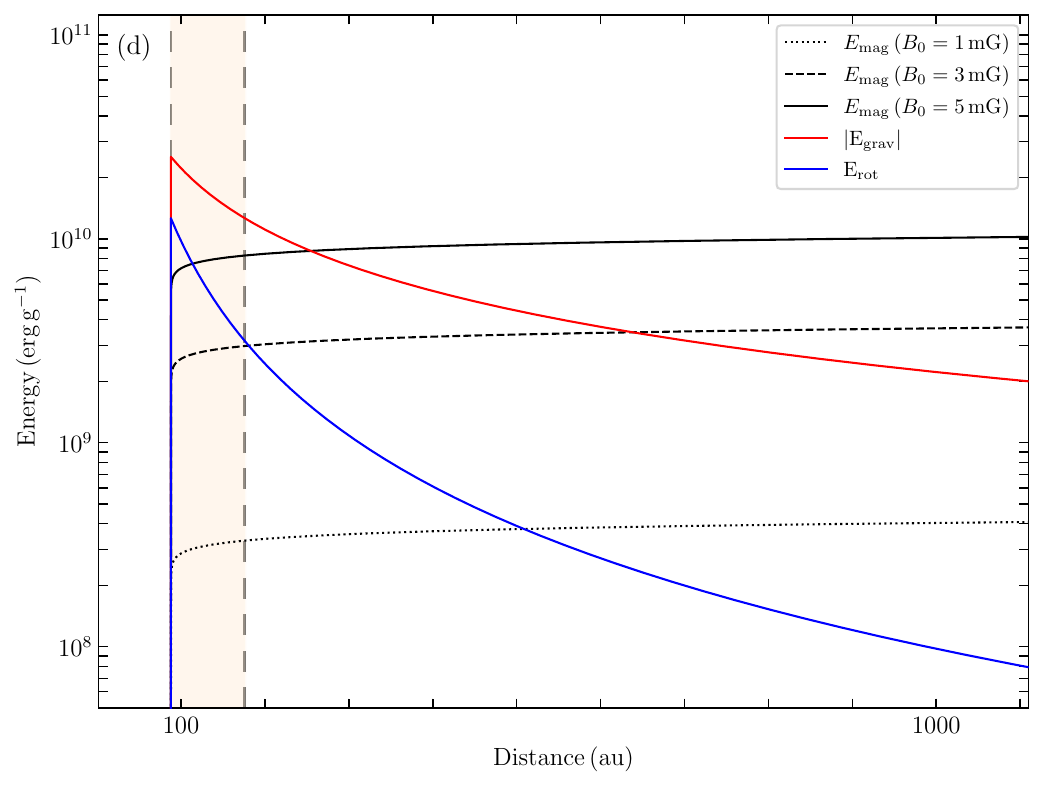}
\includegraphics[scale=0.5]{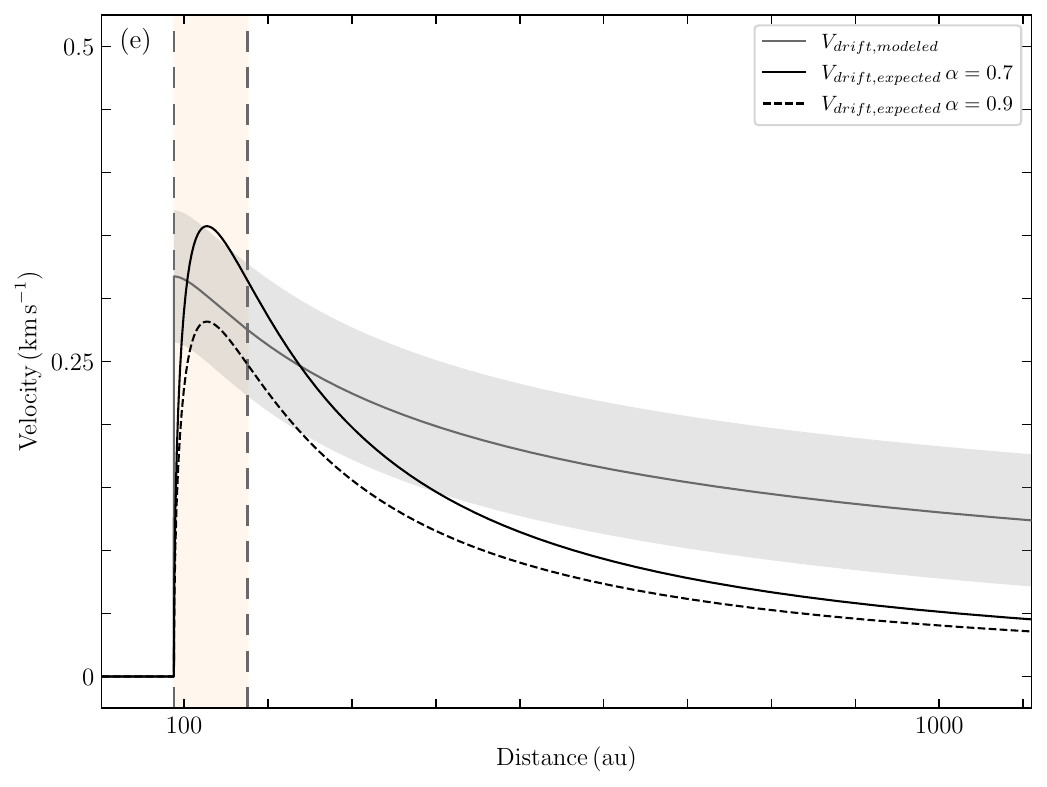}
\caption{Radial profiles of the physical properties of the envelope of HH 212. (a) Infall (solid lines) and rotational (dashed lines) velocities, with the radial velocity drift shown as a gray solid line and the sound speed indicated by a black dashed-dotted line. (b) AD fraction. (c) Hall parameter. (d) Gravitational (red line), rotational (blue line), and magnetic (black lines) energies per unit mass. (e) Radial drift velocity from the model (gray line) and expected drift velocities (black lines). The shaded region in panel (a) represents the channel width from the observations. The shaded region corresponds to the radial range between $R_{cen}\,(\sim88\,\mathrm{au})$ and $2R_{cen}\,(\sim176\,\mathrm{au})$.}
\label{fig:profiles}
\end{figure*}

\subsection{Ambipolar Diffusion Effects}
\label{subsec:ambipolar}

For the ionized tracer model (H$^{13}$CO$^{+}$) with higher ionization rate, we adopt $\alpha=0.7$, and for the neutral tracer (C$^{17}$O) we use $\alpha=0.9$. This assumption leads to a radial velocity drift of $V_{drift}\sim0.1-0.3\,\pm\,0.05$ km s$^{-1}$ along the major axis (gray line in Figure \ref{fig:profiles}a), with the range increasing toward smaller distances from the central star and becoming narrower at larger distances. The $\pm0.05$ km s$^{-1}$ represents half the channel width. This feature is consistently observed across the sampled envelope, suggesting that it is robust and not dependent on the specific region considered. The isothermal sound speed, $c_s=\sqrt{k T/ \mu m_H}\sim 0.51$ km s$^{-1}$ (\citealt{Lee2014}), is larger than the observed velocity drift, indicating that the relative motion between ions and neutrals is subsonic. This velocity difference is interpreted as being consistent with AD, an important non-ideal MHD effect in the low-density protostellar envelope, which arises from the partial decoupling of the neutral gas from the charged species that remain tightly coupled to the magnetic field (\citealt{Mestel1956}). As a result, the neutrals can drift relative to the field lines, allowing them to slip through the magnetic flux, achieving a larger infall velocity than the charged species.

The inferred ion-neutral drift velocity of $\sim0.1-0.3\pm0.05$ km s$^{-1}$ should be interpreted with caution, since it is comparable to the spectral resolution of 0.105 km s$^{-1}$. In addition, the $\alpha$ values were explored using discrete values rather than being obtained from a continuous fit with formal uncertainties. Therefore, the velocity difference between C$^{17}$O and H$^{13}$CO$^+$ should be regarded as a model-dependent estimate of the ion-neutral drift rather than a direct measurement.

This departure from ideal flux-freezing is critical: if the magnetic field were perfectly coupled to the infalling material, inward advection of field lines would dramatically enhance magnetic tension, resulting in excessively efficient magnetic braking (\citealt{Mouschovias1991}). AD mitigates magnetic braking and enables the collapsing gas to retain sufficient angular momentum to form a stable protostellar disk (\citealt{Mellon2009}). Additional non-ideal terms—most notably the Hall effect and Ohmic resistivity, become relevant in the denser inner regions but play a secondary role at the envelope scales considered here (e.g., \citealt{Nakano2002,Shu2006}).

To quantify the role of AD in the envelope, we evaluate the drift velocity between neutrals and ions. To characterize the relative strength of AD in a dimensionless form, we define the AD fraction as
\begin{equation}
    \label{eq:ADfraction}
    f_{AD}=\frac{V_{drift}}{V_{A}},
\end{equation}
where $V_{A}=B/\sqrt{4\pi\rho}$ is the Alfv\'en velocity of the gas. Here, $B$ represents the magnetic field strength, and $\rho$ is the mass density of H$_2$. For clouds where AD is considered to contribute to the collapse dynamics, the magnetic field strength scales as (\citealt{Fiedler1993})
\begin{equation}
    \label{eq:bfield}
    B=B_0 \left(\frac{\rho}{\rho_0}\right)^{0.47},
\end{equation}
where $B_0$ and $\rho_0$ correspond to the magnetic field and density at the reference radius $R_0=2R_{cen}$. Assuming a magnetic field of $B_0\sim 1-5$ mG, typical for Class 0 low-mass protostars (e.g., \citealt{Girart2006,Stephens2013,Kwon2019}), and a mass density of H$_2$ given by $\rho=1.4m_{H_2} n(H_2)$, where $n(H_2)$ is obtained from our models, we can estimate the magnetic field strength, the Alfv\'en velocity, and finally $f_{AD}$ as a function of distance from the central star, which is shown in Figure \ref{fig:profiles}b. 
For the weakest magnetic field, $B_0=1$ mG, $f_{AD}$ increases from $\sim0.4$ in the outer envelope to $\sim1$ around $2R_{cen}$, indicating that the drift velocity grows and approaches the Alfv\'en velocity, with increasing ion-neutral decoupling toward the inner envelope. For intermediate and strong magnetic fields ($B_0=3-5$ mG), $f_{AD}$ varies from $\sim0.1$ in the outer envelope to $\sim0.2$–$0.4$ toward the inner regions, indicating that AD is present, but the drift velocity remains below the Alfv\'en velocity for the assumed magnetic field strengths, consistent with a regime where magnetic forces remain dynamically relevant but are not directly constrained observationally. As the density in the midplane of the Ulrich model diverges at $R=R_{cen}$, which may lead to an overestimation of the derived quantities in the shaded region; however, these values are not used in the interpretation of our results.

The ion-neutral velocity drift suggests that the ionized molecules are decoupled from the neutral molecules. This also indicates that the ionized molecules are attached to the magnetic field throughout the envelope, with the effect being stronger near the centrifugal radius. The degree to which H$^{13}$CO$^{+}$ is tied to the magnetic field depends on the balance between the Lorentz force and the drag force acting on them, which can be quantified using the Hall parameter (\citealt{Zhao2016}) given by
\begin{equation}
    \label{eq:beta}
    \beta_{i,H_2}=\left(\frac{Z_ieB}{m_i c}\right)\frac{m_i+m_{H_2}}{\mu m_H n(H_2) \langle \sigma_i v\rangle_{i,H_2}},
\end{equation}
where $m_i$ and $Z_i e$ are the mass and charge of H$^{13}$CO$^{+}$, $c$ is the speed of light, $\mu$ is the mean molecular weight of 2.33, $m_H$ and $m_{H_2}$ are the masses of atomic and molecular hydrogen, and $\langle \sigma_i v\rangle_{i,H_2}$ is the momentum transfer rate coefficient. The coefficient $\langle \sigma_i v\rangle_{i,H_2}$ is a function of the temperature and velocity difference between ions and neutrals, with $\langle \sigma_i v\rangle_{i,H_2}\propto V_{rms}^{0.6}$ for HCO$^{+}$ (\citealt{Pinto_2008}), where $V_{rms}\equiv (V_{drift}^2+V_{th}^2)^{1/2}$ and $V_{th}=\sqrt{8kT/\pi\mu}$ is the thermal velocity, and assuming a similar dependence for H$^{13}$CO$^{+}$. For the temperature and velocity drift obtained from our models and the reduced mass appropriate for H$^{13}$CO$^{+}$-H$_2$ collisions, we estimate $\langle \sigma_i v\rangle_{i,H_2}\approx 2.0\times10^{-9}$ cm$^{3}$ s$^{-1}$. Considering the magnetic field strength defined before, the Hall parameter as a function of distance from the central star is shown in Figure \ref{fig:profiles}c. When $\beta_{i,H_2}\gg1$, meaning that the Lorentz force dominates over the drag force, the ionized molecules are attached to the magnetic field, whereas when $\beta_{i,H_2}\ll1$, the drag force dominates, implying that the ionized species are no longer well coupled to the magnetic field and tend to follow the neutral gas motion. Figure \ref{fig:profiles}c shows that $\beta_{i,H_2}\gg1$ for all considered magnetic field strengths, indicating that the ionized species remain strongly coupled to the magnetic field throughout the envelope. 

The trends observed in the AD fraction $f_{AD}$ and the Hall parameter $\beta_{i,H_2}$ in our envelope models are in agreement with theoretical predictions and numerical simulations of non-ideal MHD including AD (e.g., \citealt{Krasnopolsky2002, Mellon2009, Li2011, Zhao2016, Zhao2018}). These works show that ambipolar diffusion leads to a gradual increase in the ion–neutral drift toward smaller radii, reflecting a change in the relative importance of drift motions compared to the Alfv\'en velocity. In our models, this behavior is captured by $f_{AD}$ in Figure \ref{fig:profiles}b, which quantifies the ratio between drift and Alfv\'en velocities and varies across the envelope. In contrast, the Hall parameter remains $\beta_{i,H_2}\gg1$ for all considered magnetic field strengths in Figure \ref{fig:profiles}c, indicating that the ionized species remain strongly coupled to the magnetic field throughout the envelope. Together, these quantities describe the spatial variation of ambipolar diffusion and ion–neutral dynamics, without implying a change in the ion–magnetic coupling regime.

The interplay between ion-neutral decoupling and magnetic forces can be assessed by comparing the gravitational ($E_{grav}=-GM/r$, red line), rotational ($E_{rot}=v_\phi^2/2$, blue line), and magnetic ($E_{mag}=B^2/8\pi\rho$, black lines) energies per unit mass as a function of distance from the central star, shown in Figure \ref{fig:profiles}d. The shaded region between $R_{\rm cen}$ and $2R_{\rm cen}$ highlights the inner part of the modeled envelope, where the rotational contribution becomes increasingly important toward smaller radii, reaching $\sim50\%$ of $|E_{grav}|$ at $R_{\rm cen}$ and decreasing to $\sim25\%$ at $2R_{\rm cen}$. This behavior is expected from the Ulrich model, in which the rotational velocity increases toward smaller radii as a consequence of the conserved specific angular momentum. Beyond $2R_{\rm cen}$, the rotational contribution decreases further, reaching $\sim4\%$ at $\sim$1100 au and becoming progressively less important at larger radii. For stronger magnetic fields ($B_0 \gtrsim 3$--5 mG), the magnetic energy can exceed the gravitational energy over a large portion of the envelope, illustrating the strong dependence of the magnetic contribution on the assumed field strength. For $B_0=1$ mG, the magnetic energy is $\sim0.4\%$ of $\left|E_{grav}\right|$ at $R_{\rm cen}$ and increases to $\sim3\%$ at $2R_{\rm cen}$. At larger radii, the magnetic contribution continues to increase, reaching $\sim20\%$ in the outer envelope. Thus, while rotation is more important in the inner envelope, the magnetic contribution becomes increasingly important relative to rotation toward the outer envelope. Moreover, these results provide a self-consistent picture with the AD diagnostic discussed above, where $f_{AD}$ quantifies the relative importance of ion–neutral drift compared to the Alfv\'en velocity, while the Hall parameter remains $\beta_{i,H_2} \gg 1$ throughout the envelope, with the ionized species remaining strongly coupled to the magnetic field at all radii. Importantly, our analysis does not imply a magnetically regulated collapse in the classical sense, but rather indicates that AD and rotation can modify the infall dynamics within the gravitationally bound envelope.

The radial component of the momentum equation describes the force balance in the equatorial plane, $\theta=\pi/2$. In a magnetized collapsing envelope, this balance includes the gravitational force as well as the magnetic-pressure and magnetic-tension forces \citep{Allen2003}. Following the radial formulation adopted by \citet{Aso2015} and \citet{Sai2023}, the radial momentum equation can be written as
\begin{equation}
\label{eq:radial_mhd}
\rho(R)V_r(R)\frac{dV_r}{dR}= -\frac{GM_*\rho(R)}{R^2} -\frac{1}{2\mu_0}\frac{dB^2}{dR} +\frac{B(R)^2}{\mu_0R_{\rm curv}(R)},
\end{equation}
where the first, second, and third terms on the right-hand side represent the gravitational, magnetic-pressure, and magnetic-tension forces, respectively.

The reduction in $V_r$ can then be used to estimate the net magnetic contribution to the radial force balance. Following the approach described in Appendix C of \citet{Sai2022}, we assume $V_{r,\rm obs}=\alpha V_r$, where $\alpha$ is constant. For the Ulrich solution in the equatorial plane, the radial velocity is $V_r=-\left(GM_*/R\right)^{1/2}$, and therefore
\begin{equation}
\label{eq:dVr}
\frac{dV_{r,\rm obs}}{dR}=\alpha\frac{dV_r}{dR}\equiv \frac{\alpha}{2}\left(\frac{GM_*}{R^3}\right)^{1/2}.
\end{equation}
Thus, the inertial term becomes
\begin{equation}
\rho(R) V_{r,\rm obs}(R)\frac{dV_{r,\rm obs}}{dR}=
-\frac{\alpha^2}{2}\frac{GM_*\rho(R)}{R^2}.
\end{equation}

We define the net magnetic force as the combined contribution of magnetic pressure and magnetic tension,
\begin{equation}
\label{eq:magnetic_force_definition}
F_B= -\frac{1}{2\mu_0}\frac{dB^2}{dR} +\frac{B^2}{\mu_0R_{\rm curv}}.
\end{equation}
Using $F_G=-GM_*\rho(R)/R^2$, the radial momentum equation can then be written as
\begin{equation}
\label{eq:magnetic_force_alpha}
F_B= \left(\frac{\alpha^2}{2}-1\right)F_G.
\end{equation}
Since $F_G<0$, the magnitude of the magnetic force opposing gravity is
\begin{equation}
\label{eq:magnetic_force_magnitude}
|F_B|=\left(1-\frac{\alpha^2}{2}\right)|F_G|.
\end{equation}
For the adopted values of $\alpha=0.7$ for H$^{13}$CO$^{+}$ and $\alpha=0.9$ for C$^{17}$O, the required magnetic forces are $|F_B|=0.76|F_G|$ and $|F_B|=0.60|F_G|$, respectively. Thus, the observed reductions in $V_r$ imply that the net magnetic force would need to oppose gravity at a level of $\sim76\,\%$ and $\sim60\,\%$. These values quantify the magnetic-force requirements associated with the adopted velocity reductions, but do not uniquely determine the magnetic field strength because $F_B$ also depends on the spatial structure and geometry of the field.

The ion-neutral drift velocity can be derived from the balance between the Lorentz force acting on the charged species and the ion-neutral drag force (\citealt{Shu1983,Shu1992}). Assuming that the magnetic force is balanced by the ion-neutral drag force, $F_B=F_{drag}$, the drift velocity can be written as
\begin{equation}
\label{eq:vdrift}
V_{drift}=\frac{F_B}{\gamma_{in}\rho_i\rho},
\end{equation}
where \(\gamma_{in}\) is the ion-neutral drag coefficient, $3.5\times10^{13}$ cm$^{3}$ g$^{-1}$ s$^{-1}$ (\citealt{Draine1983}), and $\rho_i$ and $\rho$ are the ion and neutral mass densities, respectively. For the magnetic force $F_B$ defined in Equation (\ref{eq:magnetic_force_magnitude}), and $\rho_i=3\times10^{-16}\rho^{1/2}$ (\citealt{Shu1992}), the expected drift velocity for $\alpha=0.7$ is $\sim0.05-0.35$ km s$^{-1}$ and for $\alpha=0.9$ is $\sim0.05-0.25$ km s$^{-1}$, as shown in Figure \ref{fig:profiles}e, where the expected drift velocity (black lines) is consistent with the drift velocity obtained from our model in the regions close to the protostar, $R\lesssim400$ au for $\alpha=0.7$ and $R\lesssim300$ au for $\alpha=0.9$. At larger radii, $R>400$ au, the expected drift velocity for both $\alpha$ values is lower than that obtained from our model. This difference indicates that the drift velocity estimated from the radial force balance does not reproduce the radial trend obtained from our model at larger radii.

In summary, our analysis of the observations and models indicates a regime in which the gas remains coupled to the magnetic field in the outer regions, while ions and neutrals show clear signs of partial decoupling closer to the center. Observations of polarized emission capable of constraining the magnetic-field strength are needed to assess its role in shaping the dynamics of the protostar–disk system.

\section{Conclusions}
\label{sec:conclusions}

We have presented a comprehensive analysis of ALMA observations of H$^{13}$CO$^{+}$ (4--3) and C$^{17}$O (3--2) molecular line emissions from the envelope associated with the protostellar system HH 212. Our main results, based on a comparison of observations with radiative transfer models interpreted in the framework of non-ideal MHD, are as follows.

\begin{enumerate}
    \item Different abundance prescriptions for C$^{17}$O and H$^{13}$CO$^{+}$ were tested within the modeling framework, yielding consistent solutions for the envelope structure. From these models, the envelope mass derived is $M_{env}\sim0.12\,M_\odot$, while the mass infall rate toward the protostar–disk system is $\dot{M}_{env}\sim 4.4-5.7\times10^{-6}\,M_\odot$ yr$^{-1}$.
    \item Comparison between ionized (H$^{13}$CO$^{+}$ (4--3)) and neutral (C$^{17}$O (3--2)) tracers reveals a model-dependent velocity drift of $V_{drift}\sim0.1-0.3$ km s$^{-1}$. This drift is consistent with AD-like effects in the envelope, with larger values toward the inner regions. Given the spectral resolution of 0.105 km s$^{-1}$ and the discrete $\alpha$ values explored in the models, the inferred velocity drift should be regarded as a model-dependent estimate rather than as a direct measurement.
    \item The Hall parameter, $\beta_{i,H_2}$, shows that the ionized molecules remain strongly coupled to the magnetic field throughout the envelope. For plausible magnetic field strengths ($B_0 = 1-5$ mG), the gas remains influenced by magnetic fields, while non-ideal MHD effects become significant in the inner regions.
    \item The comparison of gravitational, rotational, and magnetic energies indicates that rotation becomes increasingly important toward smaller radii, reaching $\sim50\,\%$ of the gravitational energy at $R_{cen}$, but its contribution decreases at larger radii. For $B_0 = 1$ mG, the magnetic energy remains subdominant to the gravitational energy, increasing from $\sim0.4\,\%$ at $R_{cen}$ to $\sim20\,\%$ in the outer envelope, whereas stronger fields ($B_0 = 3$--5 mG) can exceed the gravitational energy over a large portion of the envelope. These results indicate that the relative importance of rotational and magnetic contributions depends strongly on radius and on the assumed magnetic field strength, while the ionized species remain strongly coupled to the magnetic field and ion–neutral drift increases toward the inner regions.
    \item The radial force-balance analysis indicates that the adopted reductions in the infall velocity require a net magnetic force opposing gravity at levels of $\sim76\,\%$ for H$^{13}$CO$^{+}$ and $\sim60\,\%$ for C$^{17}$O. These values quantify the magnetic-force contribution required by the inferred velocity reductions, but do not uniquely constrain the magnetic field strength because the magnetic force depends on the field strength, spatial structure, and geometry.
\end{enumerate}

\begin{acknowledgments}
We thank an anonymous referee for very useful suggestions that improved the presentation of this paper. J.A.L.-V., C.-F.L., and J.-H.L. acknowledge support from the National Science and Technology Council (NSTC) of Taiwan through grants 112-2112-M-001-039-MY3, 113-2124-M-001-008, 114-2124-M-001-015, and 115-2112-M-001-058-MY3; the Academia Sinica Investigator Award (AS-IA-108-M01); and the Taiwan Astronomical Research Alliance (TARA, formerly TAOvA), supported by NSTC grant 113-2740-M-008-005. TARA is committed to advancing astronomy in Taiwan and paving the way for the establishment of a national observatory. C.C. and L.P. acknowledge the PRIN-MUR 2020  BEYOND-2p (Astrochemistry beyond the second period elements, Prot. 2020AFB3FX), the project ASI-Astrobiologia 2023 MIGLIORA (Modeling Chemical Complexity, F83C23000800005), the INAF-GO 2024 fundings ICES (Tracking the history of ices from the cradles of planets to comets, CUP: C13C25000310001), and the INAF-GO 2023 fundings PROTO-SKA (Exploiting ALMA data to study planet forming disks: preparing the advent of SKA, C13C23000770005). L.P. and C.C. also acknowledge financial support under the National Recovery and Resilience Plan (NRRP), Mission 4, Component 2, Investment 1.1, Call for tender No. 104 published on 2.2.2022 by the Italian Ministry of University and Research (MUR), funded by the European Union – NextGenerationEU– Project Title 2022JC2Y93 ChemicalOrigins: linking the fossil composition of the Solar System with the chemistry of protoplanetary disks – CUP C53D23001600006 - Grant Assignment Decree No. 962 adopted on 30.06.2023 by the Italian Ministry of University and Research (MUR). J.M.J.-D. gratefully acknowledges the financial support provided by the SECIHTI through the Estancias Posdoctorales por Mexico CVU 864232 and the support from UNAM DGAPA-PAPIIT grant IG101224. This letter makes use of the following ALMA data: ADS/JAO.ALMA \#2016.1.01475.S. ALMA is a partnership of ESO (representing its member states), NSF (USA) and NINS (Japan), together with NRC (Canada), NSTC and ASIAA (Taiwan), and KASI (Republic of Korea), in cooperation with the Republic of Chile. The Joint ALMA Observatory is operated by ESO, AUI/NRAO and NAOJ.
\end{acknowledgments}

\bibliography{sample7}{}
\bibliographystyle{aasjournal}

\end{document}